**This is the accepted manuscript (postprint) of the following article:**
R. Vahedifard, E. Salahinejad, *Microscopic and spectroscopic evidences for multiple ion-exchange reactions controlling biomineralization of CaO. MgO. 2SiO2 nanoceramics*, Ceramics International, 43 (2017) 8502-8508.
https://doi.org/10.1016/j.ceramint.2017.04.001

# Microscopic and spectroscopic evidences for multiple ion-exchange reactions controlling biomineralization of $CaO.MgO.2SiO_2$ nanoceramics

R. Vahedifard, E. Salahinejad *

Faculty of Materials Science and Engineering, K.N. Toosi University of Technology, Tehran, Iran

## Abstract

This study is focused on the mechanism of *in vitro* biomineralization on the surface of $CaO.MgO.2SiO_2$ (diopside) nanostructured coatings by scanning electron microscopy, energy-dispersive X-ray spectroscopy and inductively coupled plasma spectroscopy assessments. A homogeneous diopside coating of almost 2 μm in thickness was deposited on a medical-grade stainless steel by coprecipitation, dipping and sintering sequences. After soaking the sample in a simulated body fluid (SBF) for 14 days, a layer with the thickness of 8 μm is recognized to be substituted for the primary diopside deposit, suggesting the mineralization of apatite on the surface. Investigations revealed that the newly-formed layer is predominantly composed of Ca, P and Si, albeit with a biased accumulations of P and Si towards the surface and substrate, respectively. The variations in the ionic composition and pH of the SBF due to the incubation of the sample were also correlated with the above-interpreted biomineralization. In conclusion, the multiple ion-exchange reactions related to Ca, Mg, Si and P were found to be responsible for the *in vitro* bioactivity of nanodiopside.

* Corresponding Author:
Email Addresses: <salahinejad@kntu.ac.ir>, <erfan.salahinejad@gmail.com>

## 1. Introduction

With the criterion of the biomaterial/tissue interface type, bioceramics are classically classified into three groups: bioinert, bioactive and bioresorbable. Regarding the last two families, surface-reactivity is known to be proportional to biodegradability in most cases. Bioactivity or surface-activity is defined as the ability to form an apatite layer on the surface, giving rise to a chemical bonding to adjacent tissues [1]. The $SiO_2$-CaO-MgO system is one of the new classes of bioceramics which is currently under investigation due to its suitable mechanical properties and osteoinductivity [2, 3]. As a member of this group, diopside ($CaO.MgO.2SiO_2$ or $CaMgSi_2O_6$) is dissolved and then replaced with bone, where the rest preserves a suitable bonding to the new bone [2, 4].

Among the different methods used to synthesize ceramics, wet-chemical approaches are promising for medical application due to purity and nanosizing. Coprecipitation is one of these wet-chemical synthesis routes of ceramic nanoparticles. In this method, organic or inorganic precursors dissolved in an aqueous or alcoholic solvent are precipitated via the addition of an alkaline agent [5]. In most cases, a following calcination process is required to obtain desirable crystalline phases. Typically, the need for a calcination temperature of 1100 °C has been reported to develop a homogeneous diopside structure, when alkoxide and nitrate precursors were used in coprecipitation [6]. Nonetheless, it has been shown that by using chloride precursors in this synthesis method, the required calcination temperature is reduced to 700 °C [7, 8]. Also, the preparation of other multi-oxides via wet-chemical synthesis routes using chloride precursors has been pointed out [9-17].

To the best of our knowledge, there is little systematic report on the study of the degradation and mineralization of nanodiopside coatings. In this work, a precipitation-derived, spin-coated, sintered diopside coating was studied from the viewpoints of structure, bioactivity and biodegradation to establish a comprehensive mechanism explaining the investigated phenomena.

## 2. Experimental procedure

In this work, creamy precipitates were synthesized by an inorganic-salt coprecipitation route using chloride precursors with the molar ratio of Ca:Mg:Si = 1:1:2 and the ammonia solution precipitant in an ethanolic solvent, according to Refs. [7, 8]. A part of the synthesized product was analyzed by thermo-gravimetry (TG), X-ray diffraction (XRD, Co Kα radiation) and transmission electron microscopy (TEM, 200 kV), where the two recent tests were done after calcination at 700 °C. Also, stainless steel 316L substrates were covered by dipping in a suspension of the washed precipitates and sintering at 700 °C. The surface and cross section of the fired coating were studied by a scanning electron microscope (SEM). Moreover, followed by soaking the samples in the simulated body fluid (SBF) [18] for a period of 14 days, the surface and cross-section were evaluated by SEM equipped with point, linear and XMAP energy-dispersive X-ray spectroscopy (EDS). The ionic concentration of the SBF after incubating the sample was also assessed by inductively coupled plasma spectroscopy (ICP), as well as pH measurements.

## 3. Results and discussion

Fig. 1 presents the TG profile of the coprecipitation-derived products. In this curve, a sharp weight loss by 65 % to 350 °C is observed due to the evaporation of adsorbed ethanol and water and the sublimation of ammonium chloride created during the wet-chemical synthesis. The loss is followed by a slow tend to 1000 °C, suggesting that the material after calcination at temperatures above 350 °C should be relatively residual-free. However, in this work, the temperature of 700 °C was selected for sintering of the deposited sol to ensure the removal of residuals and the development of a homogeneous structure of diopside ($CaMgSi_2O_6$) on the surface, as inferred from the XRD analysis of the related powder (Fig. 2). The crystallite size is also estimated to be almost 50 nm, calculated by the Scherrer equation on the most intense XRD peak, whereas the average particle size was 70 nm (Fig. 3).

The SEM micrographs of the sintered coating surface are indicated in Fig. 4 in tow magnifications. As can be seen, the coating exhibits a crack-free, relatively-uniform and nanoporous appearance, where its mean particle size is almost 140 nm. This porous and fine structure is promising for biomineralization via offering a high surface area and reactivity. Based on Fig. 5, the coating also reveals a relatively uniform thickness of about 2 μm, with a suitable adhesion to the substrate via by an oxygen-shared mechanism [19, 20]. In this regard, the sintering temperature of 700 °C has been recognized to be optimal for oxide coatings deposited on stainless steels [21].

Fig. 6 represents the SEM micrographs of the coating surface after soaking in the SBF for 14 days. A comparison between Figs. 4 and 65 suggests the precipitation of apatite on the surface due to immersion. According to Fig. 6a, the apatite precipitates have completely covered the surface, but in two different morphologies: a bottom integrated layer of intersecting irregularly-oriented blades and a top non-integrated layer of the rose-like morphology. The average size and distance of the roses are almost 1 and 3 μm, respectively. Also, the largest and smallest dimensions of the apatite plates are about 180 and 25 nm, respectively, according to the high-magnification micrographs shown in Fig. 6b.

The cross-sectional backscattered-electron SEM micrograph of the sample immersed in the SBF after mounting in an epoxy resin is shown in Fig. 7a. Note that the backscattered-electron mode is mainly sensitive to chemical composition, rather than roughness. Similar to the cross section before immersion (Fig. 5), there are three different contrasts in the micrograph of the soaked sample, where they can be the substrate, coating and mount from bottom to top. However, the typical point is that the thickness of the coating layer has reached 8 μm after soaking in the SBF, i.e. quadruple the diopside coating's thickness. Also, this 6-μm added thickness exhibits no significant contrast with the initial layer, suggesting a relatively homogenous composition over the overall 8-μm thickness.

The linear EDS analysis related to the SEM micrograph represented in Fig. 7a from point A to point D is also depicted in Fig. 7b. Considering both Figs. 7a and 7b together, it is concluded that the top layer is the mount and the bottom layer is the stainless steel substrate. More importantly, the middle layer is a relatively homogenous layer of Si-Ca-Mg-P-O in thickness, but with a biased accumulation of Ca and P towards the surface and that of Si towards the coating/substrate. This confirmed the precipitation of apatite (calcium phosphate) on the surface due to incubation in the SBF, implying the apatite-forming ability of nanodiopside. The presence of P in the Si-rich layer is also noticeable, suggesting the diffusion of this species from the SBF into this layer. The aforementioned biases are also verified by the point EDS analyses taken of points A, B, C and D (Figs. 7c, 7d, 7e and 7f, respectively). In this regard, point A is C-rich (mount), point B is P-Ca-rich (apatite), point C is Si-rich (a silica layer albeit containing P and Ca), and point D is Fe-rich (substrate). Despite the considerable presence of Mg and Si in stoichiometric diopside, the absence of the former in the above-characterized layers and the reduced concentration of the latter in the majority of the interlayer are indicative of their dissolution from the diopside coating into the SBF. On the other hand, considering the absence of P in stoichiometric diopside, the presence of this species in the deposited interlayer is a result of the enrichment of P from the SBF into the surface. These two recent statements infer the multiple ion-exchange reactions accompanied by the formation of apatite on the diopside's surface due to immersion in the SBF. The Ca species also experiences a similar type of this exchange reaction, but since this ion exists in both diopside and apatite, its exchange between the SBF and surface is not straightforwardly recognizable. The XMAPs (Fig. 8) also verify the conclusion drawn above, inferring the formation a mixed layer of silica and apatite, substituting for diopside, as a result of soaking in the SBF, with a biased concentration of P and Si at the surface and substrate, respectively.

The variation in the ionic composition of the SBF due to the inclusion of the sample for 14 days is demonstrated in Fig. 9. As can be seen, the concentrations of Si, Mg and Ca have been increased, showing the dissolution of these species. In contrast, P ions have been exposed to a decline in level due to the apatite precipitation, as realized from Figs. 6, 7 and 8. These decreasing and increasing trends in the SBF suggest the multiple ion-exchange reactions during the incubation of the sample in the SBF, confirming the above SEM-EDS assessments. The pH value of the SBF was also changed from 7.4 for the fresh state to 7.8 after immersion, which is attributed to the ion-exchange reactions between the SBF and surface. On the one hand, the dissolution of Mg, Ca and Si cations from diopside into the SBF, as revealed in Fig. 9, creates a number of vacancies with a dominant tendency to absorb $H^+$ from the SBF, thereby increasing pH. On the other hand, the precipitation of apatite requires the consumption of $OH^-$ in the SBF, providing a tendency towards decreasing the pH value of the SBF. Thus, from the increase of pH measured due to soaking the sample, it is concluded that the dissolution of the cations prevails over the apatite mineralization.

In conclusion, the above-interpreted microscopic and spectroscopic evaluations infer the multiple ion-exchange reactions related to Ca, Mg, Si and P which control the *in vitro* biomineralization of nanodiopside. It is worth to mention that a similar mechanism has been recognized for the signification biomineralization of bioactive glass-ceramics and glasses [22-24]. On the contrary, crystalline Ca-free ceramics are exposed to a hydrophilicity-based mechanism for their less bioactive [25, 26].

## 4. Conclusion

In this research, the *in vitro* bioactivity of diopside coatings on stainless steel 316L was studied. The following conclusions can be drawn from the work:

a) Nanostructured single-phase diopside coatings were successfully deposited on the substrate with no crack, a proper adhesion and a thickness of 2 µm.

b) A mixed layer of Ca, P and Si with thickness of 8 µm was replaced with the primary diopside coating due to incubation in the SBF.

c) In the biomineralized layer, a biased accumulation of P and Si towards the surface and substrate was recognized.

d) The variation of the SBF composition was explained by the degradation of diopside and the precipitation of apatite.

e) Considering both the diopside's degradation and biomineralization together, a multiple ion-exchange mechanism was explored to control the apatite-forming ability.

## Figures

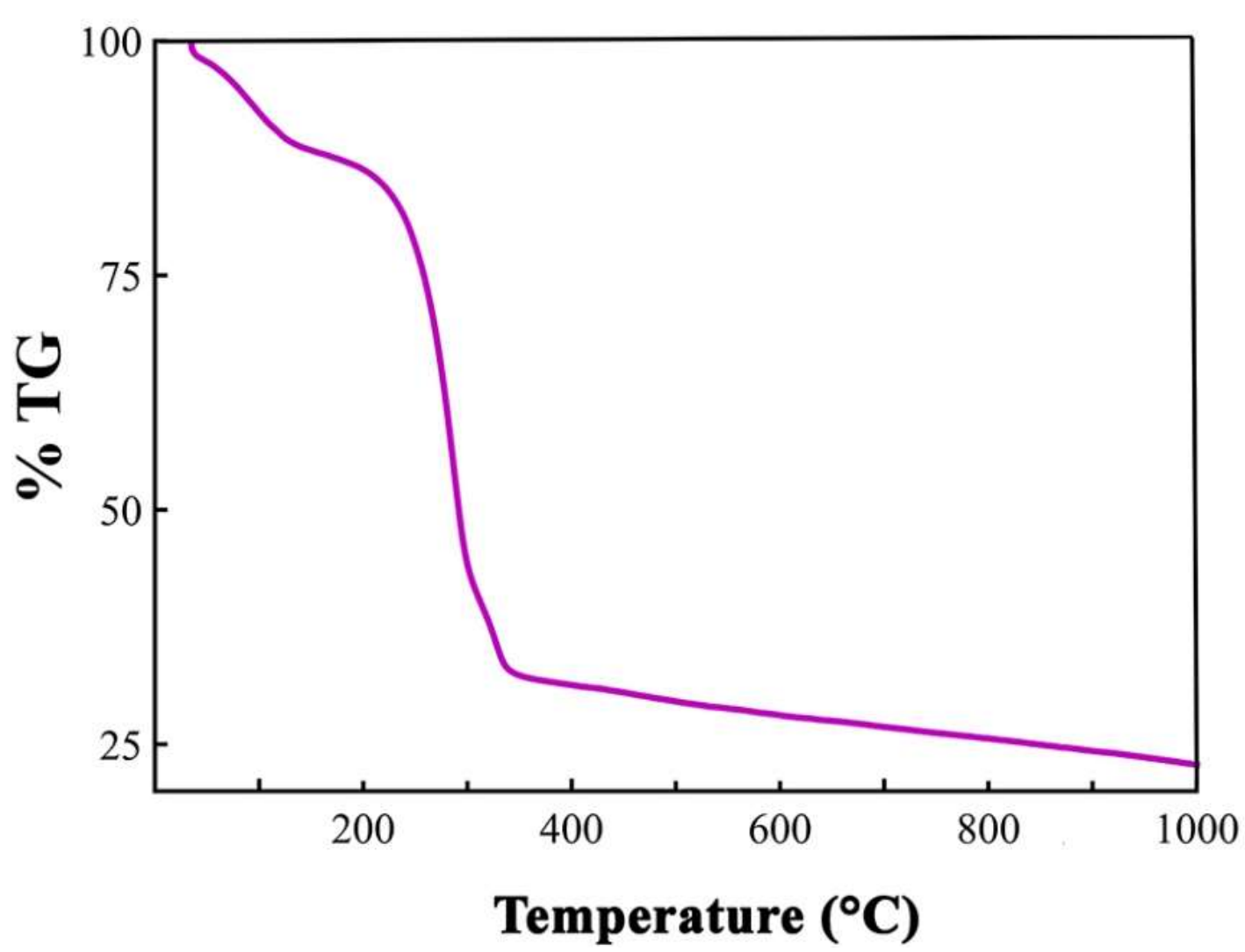

Fig. 1. TG profiles of the synthesized precipitates.

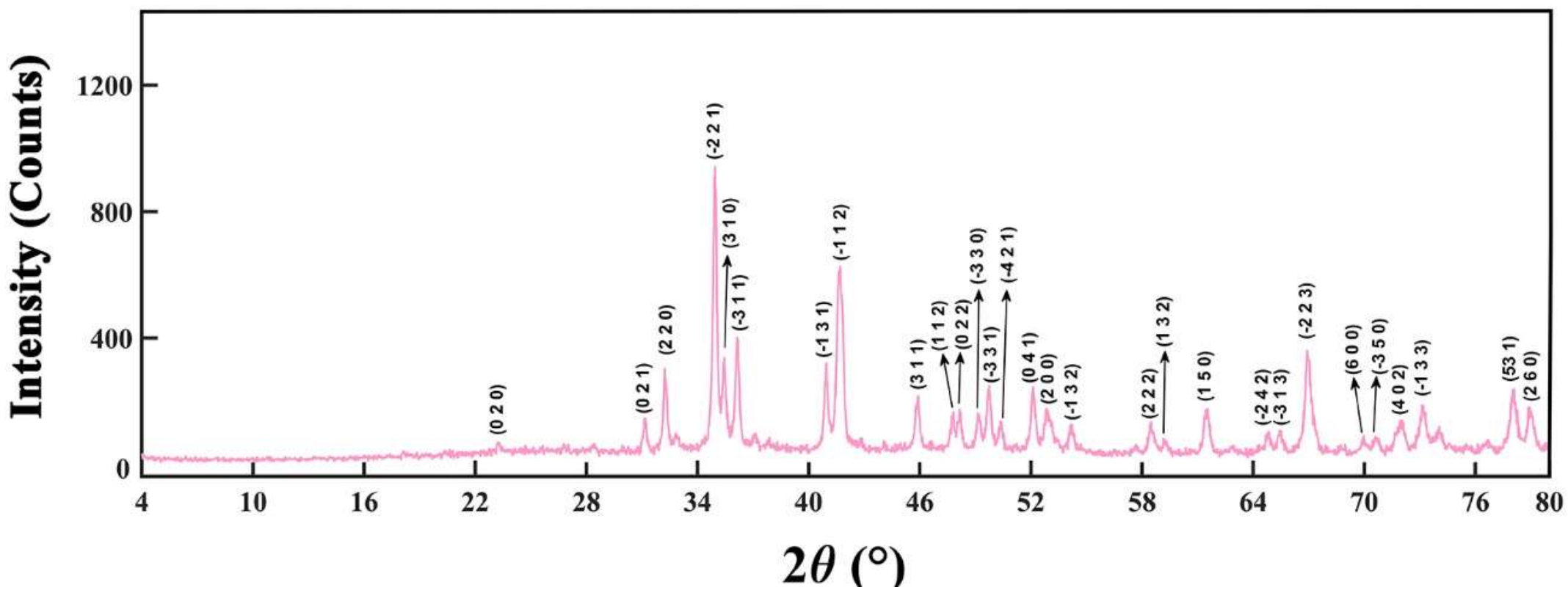

Fig. 2. XRD patterns of the powder sample calcined at 700 °C.

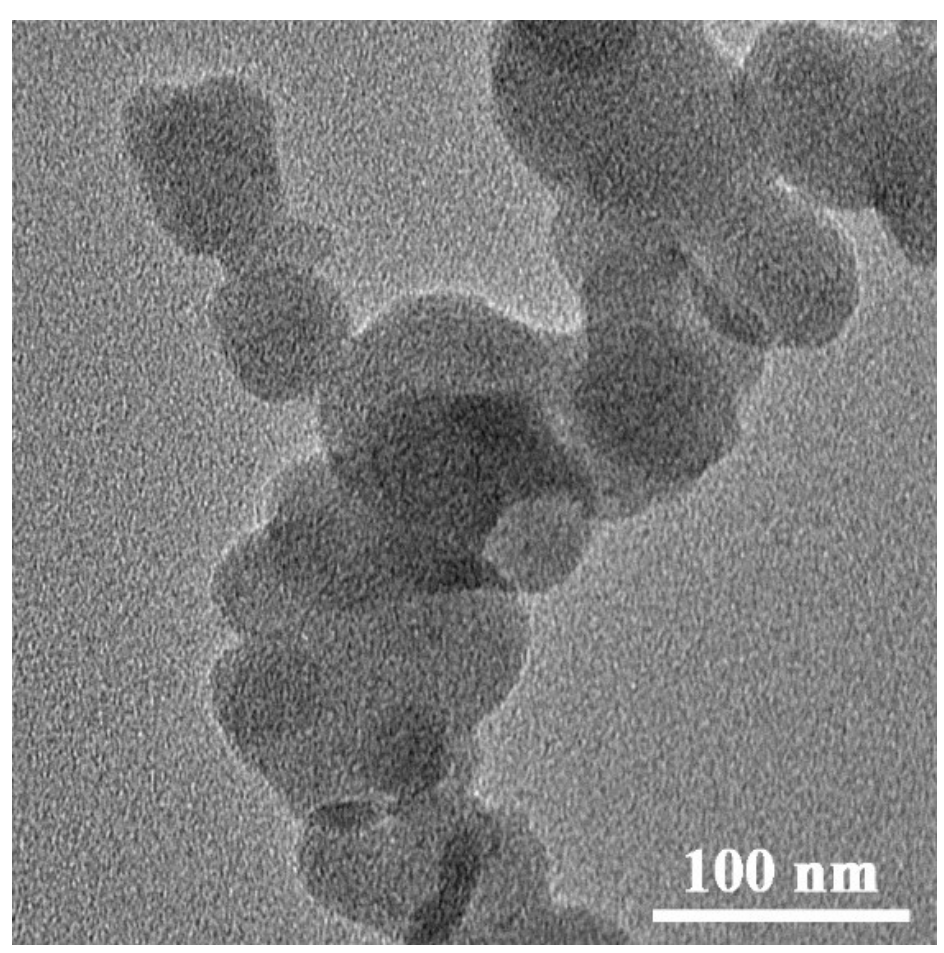

Fig. 3. TEM micrograph of the powder calcined at 700 °C.

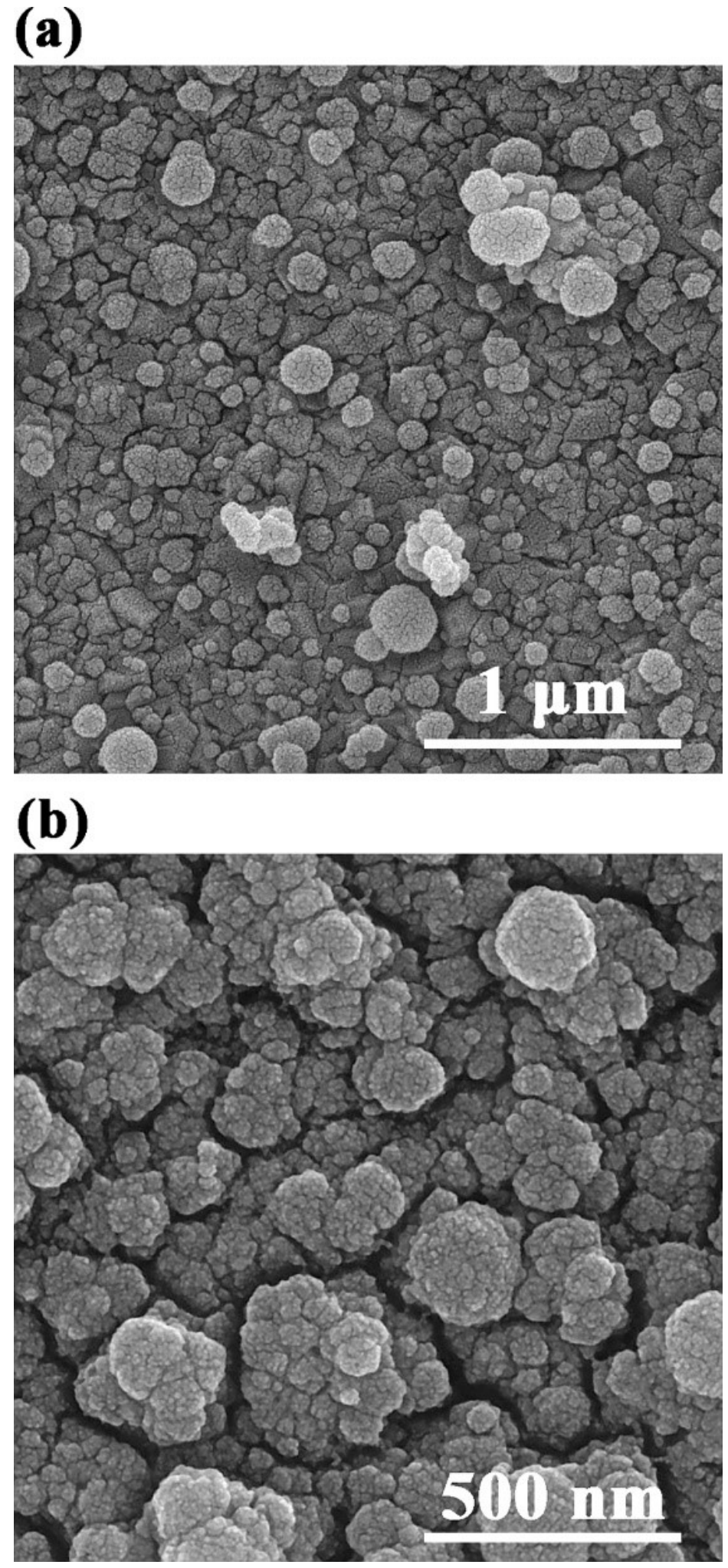

Fig. 4. Low-magnification (a) and high-magnification (b) top-view SEM micrographs of the coated sample before incubation.

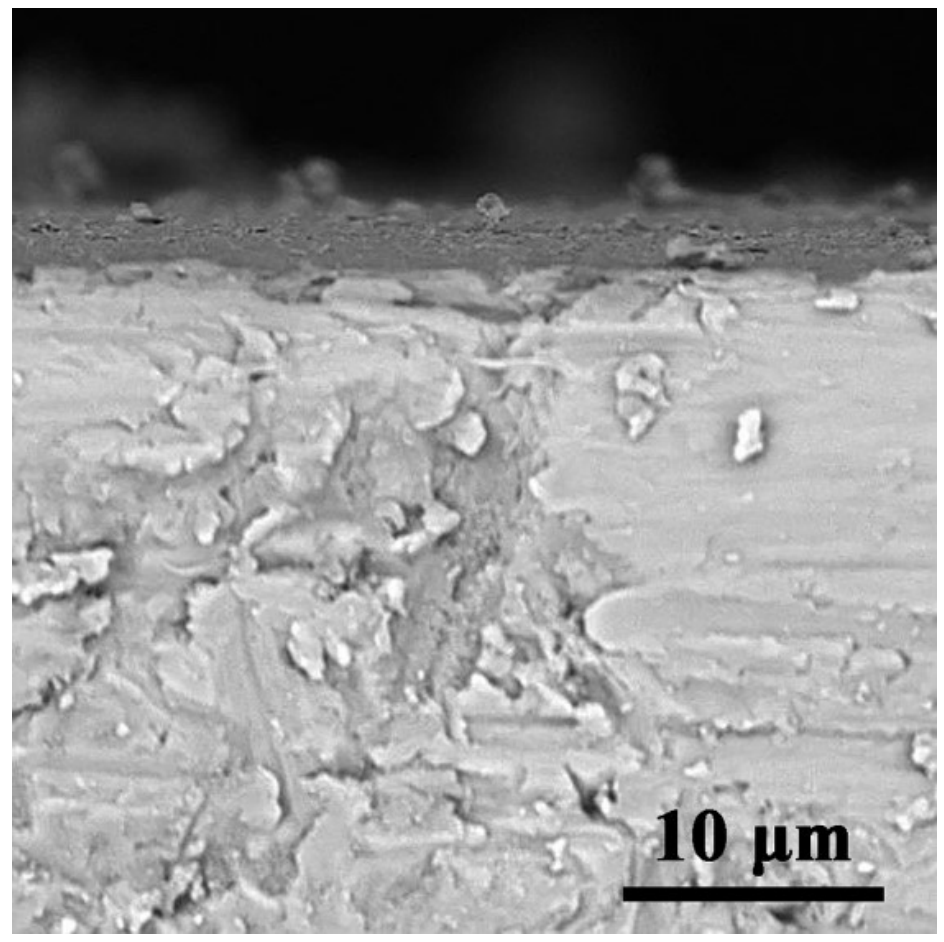

Fig. 5. Cross-sectional SEM micrograph of the coated sample before incubation.

**(a)**

**(b)**

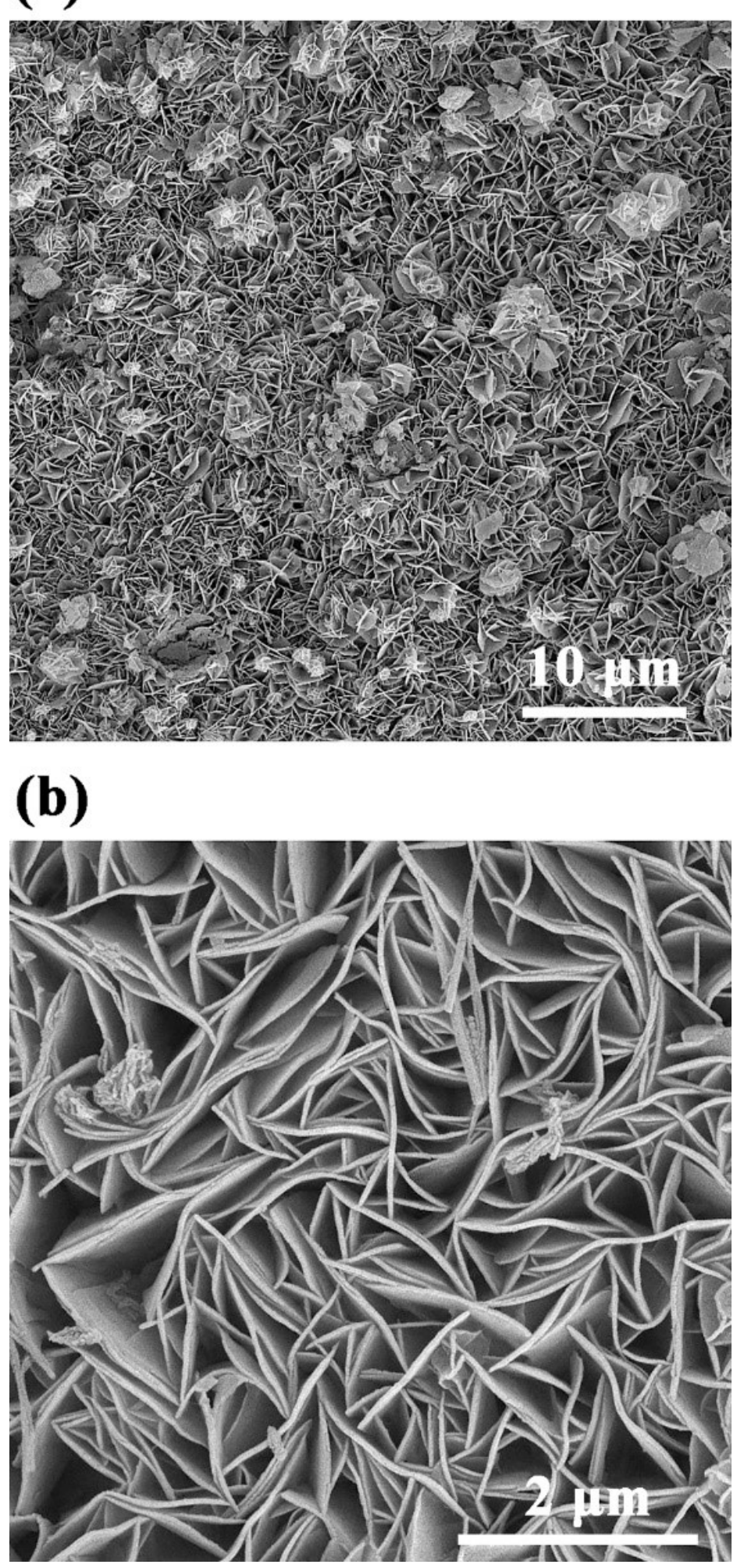

Fig. 6. Low-magnification (a) and high-magnification (b) top-view SEM micrographs of the soaked sample.

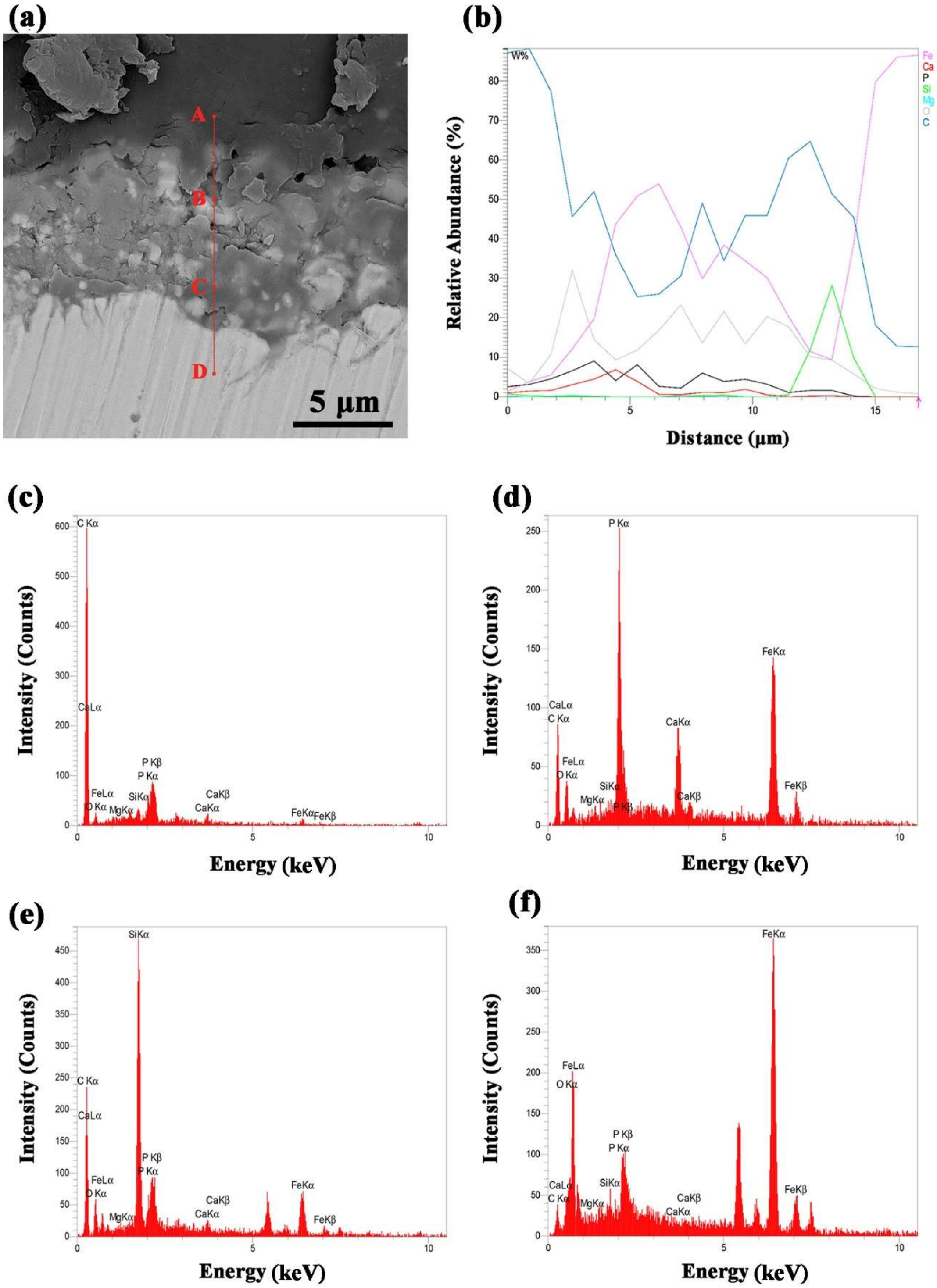

Fig. 7. Cross-sectional SEM micrograph of the soaked sample (a), linear EDS scan along the line drawn in the related micrograph (b) and point EDS spectra of points A (c), B (d), C (e) and D (f) indicated in the micrograph.

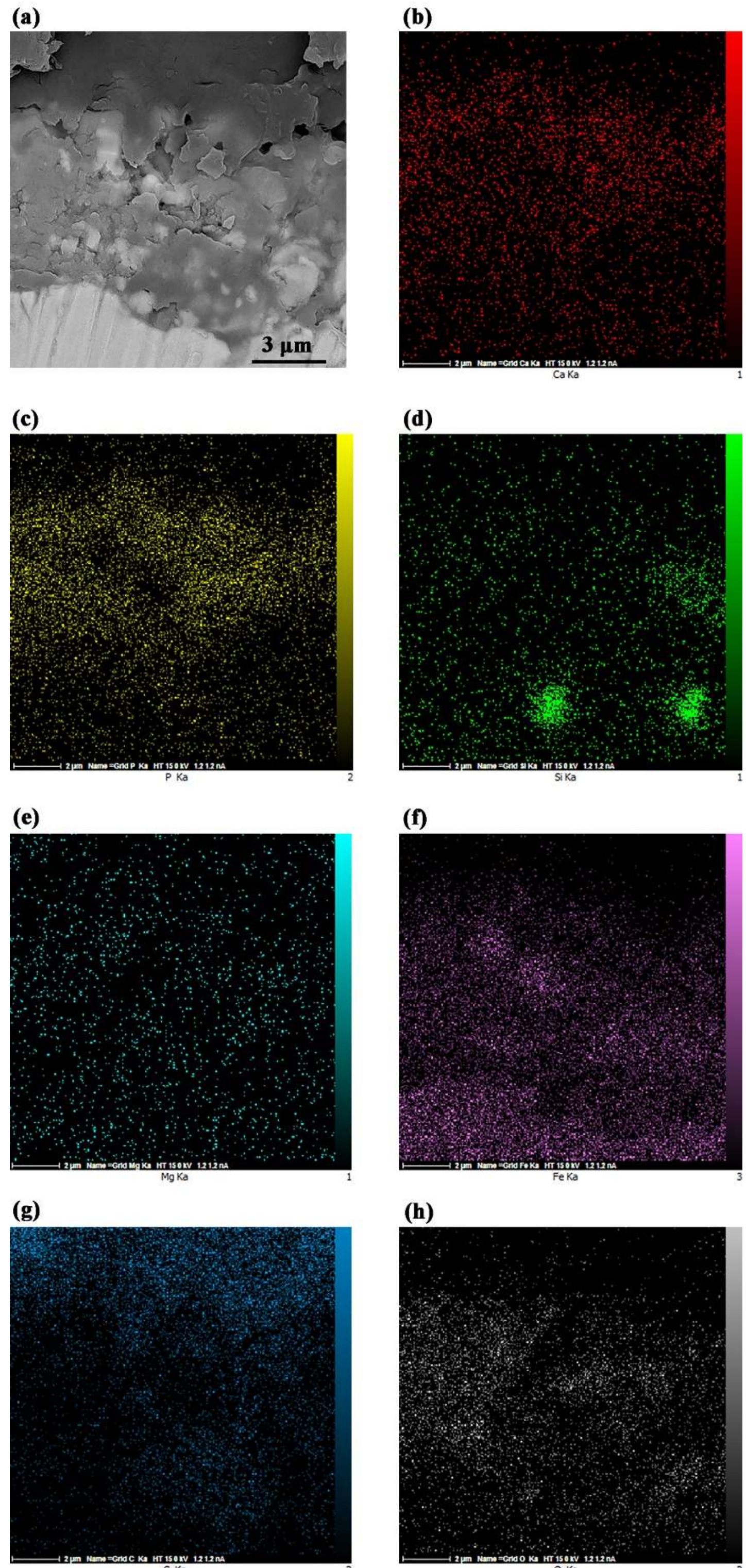
(a)
3 µm
(b)
Ca Ka
(c)
P Ka
(d)
Si Ka
(e)
Mg Ka
(f)
Fe Ka
(g)
C Ka
(h)
O Ka

Fig. 8. Cross-sectional SEM micrograph of the soaked sample (a) and related XMAP for Ca (b), P (c), Si (d), Mg (e), Fe (f), C (g) and O (h).

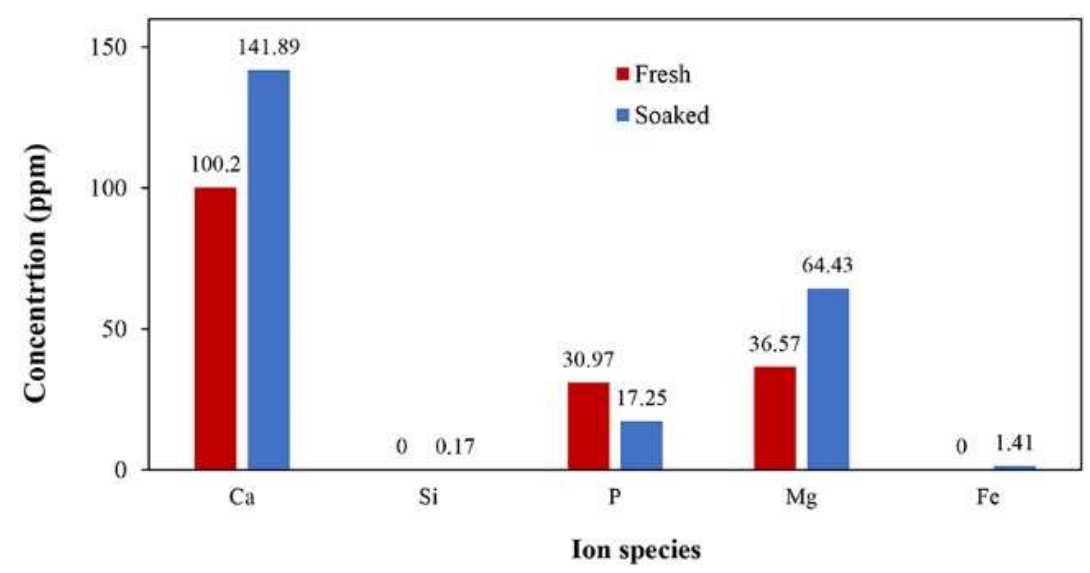

Fig. 9. Comparison of the SBF composition before and after the sample incubation.